# Cortical current source connectivity by means of partial coherence fields

Roberto D. Pascual-Marqui[1,2], Rolando J. Biscay[3], Pedro A. Valdes-Sosa[4], Jorge Bosch-Bayard[5], Jorge J. Riera-Diaz[6]

1: The KEY Institute for Brain-Mind Research, University Hospital of Psychiatry, Zurich, Switzerland (pascualm@key.uzh.ch)
2: Department of Neuropsychiatry, Kansai Medical University Hospital, Osaka, Japan (pascualr@takii.kmu.ac.jp)
3: DEUV-CIMFAV, Facultad de Ciencias, Universidad de Valparaiso, Chile (rolando.biscay@uv.cl)
4: Cuban Neuroscience Center, Havana, Cuba (peter@cneuro.edu.cu)
5: Cuban Neuroscience Center, Havana, Cuba (bosch@cneuro.edu.cu)
6: Institute of Development, Aging and Cancer; Tohoku University, Sendai, Japan (riera@idac.tohoku.ac.jp)

## 1. Abstract

An important field of research in functional neuroimaging is the discovery of integrated, distributed brain systems and networks, whose different regions need to work in unison for normal functioning.

The EEG is a non-invasive technique that can provide information for massive connectivity analyses. Cortical signals of time varying electric neuronal activity can be estimated from the EEG. Although such techniques have very high time resolution, two cortical signals even at distant locations will appear to be highly similar due to the low spatial resolution nature of the EEG.

In this study a method for eliminating the effect of common sources due to low spatial resolution is presented. It is based on an efficient estimation of the whole-cortex partial coherence matrix. Using as a starting point any linear EEG tomography that satisfies the EEG forward equation, it is shown that the generalized partial coherences for the cortical grey matter current density time series are invariant to the selected tomography. It is empirically shown with simulation experiments that the generalized partial coherences have higher spatial resolution than the classical coherences. The results demonstrate that with as little as 19 electrodes, lag-connected brain regions can often be missed and misplaced even with lagged coherence measures, while the new method detects and localizes correctly the connected regions using the lagged partial coherences.

## 2. Introduction

In its early development, methods of analysis in functional neuroimaging were aimed at the "localization" of "effects". For instance, by comparing normal control subjects and patients suffering schizophrenia ("effect"), a small number of brain regions ("localization") were claimed to be responsible for the disorder. The field has evolved and moved on, and it is now more frequent to find methods aimed at the discovery of integrated distributed systems and networks, whose different brain regions need to work in unison for normal functioning. An example of such methods used for this purpose is based on the analysis of massive amounts of pairwise similarity measures between cortical signals, i.e. on the analysis of the extremely high dimensional similarity matrix given by correlations or coherences.

Extensive general reviews on these types of methods can be found in Valdes-Sosa et al (2011) and Sporns (2011).

In the case of metabolic functional neuroimaging methods such as fMRI and PET, there is high spatial resolution, but low time resolution, with signals having most of their spectral power concentrated at frequencies lower than 0.1 Hz. In the case of EEG-based neuroimaging, the spatial resolution is lower, but with very high time resolution, with effective sampling rates typically higher than 100 Hz.

The problem of interest in this study consists of estimating intracortical connectivities from computed signals of electric neuronal activity, obtained from non-invasive scalp EEG recordings. The correlation or coherence matrices have very high dimension, equal to the number of cortical grey matter voxels at which the current density is computed (typically more than 6000), but these matrices are of low rank, due to the small number of scalp electrodes (typically ranging from 19 to 128). Due to the low spatial resolution nature of linear, discrete, distributed EEG inverse solutions, the similarity between pairs of computed cortical signals will be much higher than the true value.

The solution presented here consists of computing partial coherences, which by themselves are of great interest because they provide information on non-mediated direct connections. In addition, partial coherences would interpret the low spatial resolution effect as "common sources", thus decreasing this effect in the estimated connectivities. The whole cortex partial coherence can be obtained from the inverse of the coherence matrix. However, since this is a very low rank matrix, the inverse does not exist. In its place a generalized inverse coherence matrix is computed, thus producing the generalized whole cortex partial coherence, endowed with higher spatial resolution. Very simple and efficient equations for the whole-cortex partial coherence are derived.

## 3. Method

### 3.1. Basic equations and definitions

Let $\boldsymbol{\Phi}_{t,i} \in \mathbb{R}^{N_E \times 1}$ denote the time domain EEG, for $N_E$ electrodes, $N_T$ discrete time samples (with $t = 1...N_T$), for $N_S$ epochs (with $i = 1...N_S$).

The EEG forward equation is:

**Eq. 1** $\quad \boldsymbol{\Phi}_{t,i} = \mathbf{K}\mathbf{J}_{t,i}$

where $\mathbf{J}_{t,i} \in \mathbb{R}^{N_V \times 1}$ is the current density, $\mathbf{K} \in \mathbb{R}^{N_E \times N_V}$ denotes the lead field matrix, and $N_V$ is the number of cortical grey matter voxels. Typically, the number of voxels is much larger than the number of scalp measurements, i.e. $N_V \gg N_E$. And without loss of generality, it will be assumed that the lead field matrix is of full row rank, i.e. for the typical situation when $N_V \gg N_E$, $rank(\mathbf{K}) = N_E$.

Without loss of generality, the derivations presented here refer to EEG, but they equivalently are valid for MEG.

Any time domain linear transform, such as the discrete Fourier transform, can be applied to Eq. 1, giving:

**Eq. 2** $\quad \mathbf{\Phi}_{\omega,i} = \mathbf{K} \mathbf{J}_{\omega,i}$

where $\mathbf{\Phi}_{\omega,i} \in \mathbb{C}^{N_E \times 1}$ and $\mathbf{J}_{\omega,i} \in \mathbb{C}^{N_V \times 1}$ denote the corresponding discrete Fourier transforms at the discrete frequency $\omega$. Other transforms can be accommodated, such as wavelets or time varying Fourier transforms.

The generalized linear inverse solution is:

**Eq. 3** $\quad \hat{\mathbf{J}}_{\omega,i} = \mathbf{T} \mathbf{\Phi}_{\omega,i}$

for any matrix $\mathbf{T} \in \mathbb{R}^{N_V \times N_E}$ that satisfies:

**Eq. 4** $\quad \mathbf{KT} = \mathbf{I}$

where $\mathbf{I}$ is the identity matrix. This is equivalent to:

**Eq. 5** $\quad \mathbf{KTK} = \mathbf{K}$

due to the fact that $\mathbf{K}$ is of full row rank (see e.g. Lutkepohl 1996, pp. 32, property 3g).

It is of the essence to note that in general, $\hat{\mathbf{J}}_{\omega,i}$ is a solution to Eq. 2, as can be confirmed by plugging Eq. 3 into Eq. 2 and making use of Eq. 4:

**Eq. 6** $\quad \mathbf{\Phi}_{\omega,i} = \mathbf{K} \hat{\mathbf{J}}_{\omega,i} = \mathbf{K} \left( \mathbf{T} \mathbf{\Phi}_{\omega,i} \right) \equiv \mathbf{\Phi}_{\omega,i}$

From Eq. 3, the Hermitian covariance matrix, which is proportional to the cross-spectral density matrix, satisfies:

**Eq. 7** $\quad \mathbf{S}_{\hat{\mathbf{j}}\omega} = \mathbf{T} \mathbf{S}_{\mathbf{\Phi}\omega} \mathbf{T}^T$

with:

**Eq. 8** $\quad \mathbf{S}_{\mathbf{\Phi}\omega} = \frac{1}{N_S} \sum_{i=1}^{N_S} \mathbf{\Phi}_{\omega,i} \mathbf{\Phi}_{\omega,i}^*$

where $\mathbf{S}_{\mathbf{\Phi}\omega} \in \mathbb{C}^{N_E \times N_E}$ is the measurement-based scalp EEG Hermitian covariance, and $\mathbf{S}_{\hat{\mathbf{j}}\omega} \in \mathbb{C}^{N_V \times N_V}$ is the estimated current density Hermitian covariance.

In general, the superscript "$T$" denotes transpose, and the superscript "*" denotes complex conjugate and transpose (as used e.g. in Eq. 7 and Eq. 8, respectively).

It is important to emphasize that the source covariance estimator $\mathbf{S}_{\hat{\mathbf{j}}\omega}$ in Eq. 7 depends on the choice of inverse matrix $\mathbf{T}$ and on the measurements via Eq. 8. This is to be contrasted with the particular estimator for the inverse source covariance below, which is independent of any linear inverse used (i.e. independent of the choice of $\mathbf{T}$).

The estimated intracranial coherences are obtained from the current density covariance in Eq. 7 as:

**Eq. 9** $\quad \mathbf{R}_{\hat{\mathbf{j}}\omega} = \mathbf{D}_{\hat{\mathbf{j}}\omega} \mathbf{S}_{\hat{\mathbf{j}}\omega} \mathbf{D}_{\hat{\mathbf{j}}\omega}$

with:

**Eq. 10** $\quad \mathbf{D}_{\hat{\mathbf{j}}\omega} = \left[ diag\left( \mathbf{S}_{\hat{\mathbf{j}}\omega} \right) \right]^{-1/2}$

where the "*diag*" operator returns a diagonal matrix (off-diagonal elements set to zero). Note that Eq. 10 is valid only if the diagonal elements:

**Eq. 11** $\left[\mathbf{S}_{\hat{\mathbf{j}}\omega}\right]_{ii} > 0 \quad for \quad i = 1...N_V$

It is of interest to minimize the effect of common sources when analyzing connectivity. This can be achieved by computing the partial coherence matrix. In practice, the computations use the well-known fact that the inverse covariance matrix corresponds to the matrix of partial covariances. From the matrix of partial covariances, the partial coherence matrix is obtained by scaling to partial covariances with the corresponding partial standard deviations.

In simple matrix algebra terms, if **S** denotes a positive definite covariance matrix, and:

**Eq. 12** $\mathbf{Q} = \mathbf{S}^{-1}$

denotes its inverse, then the partial coherence is:

**Eq. 13** $\mathbf{P} = \mathbf{EQE}$

with:

**Eq. 14** $\mathbf{E} = \left[diag(\mathbf{Q})\right]^{-1/2}$

This definition can be generalized to the case of a non-negative definite covariance matrix, such as the estimated current density covariance in Eq. 7, which is singular, with $rank(\mathbf{S}_{\hat{\mathbf{j}}\omega}) = N_E \ll N_V$.

### 3.2. The inverse source covariance

We propose as an estimator for the inverse source covariance the particular form:

**Eq. 15** $\mathbf{S}_{\hat{\mathbf{j}}\omega}^{-r} = \left(\mathbf{TS}_{\Phi\omega}\mathbf{T}^T\right)^{-r} = \mathbf{K}^T\mathbf{S}_{\Phi\omega}^{-1}\mathbf{K}$

where the superscript "-r" denotes a reflexive generalized inverse. Note that this generalized inverse does indeed satisfy the two properties that characterize a reflexive inverse:

**Eq. 16** $\mathbf{S}_{\hat{\mathbf{j}}\omega}\mathbf{S}_{\hat{\mathbf{j}}\omega}^{-r}\mathbf{S}_{\hat{\mathbf{j}}\omega} = \mathbf{S}_{\hat{\mathbf{j}}\omega} \quad \left[based \; on \; \mathbf{TS}_{\Phi\omega}\mathbf{T}^T\mathbf{K}^T\mathbf{S}_{\Phi\omega}^+\mathbf{KTS}_{\Phi\omega}\mathbf{T}^T = \mathbf{TS}_{\Phi\omega}\mathbf{T}^T\right]$

**Eq. 17** $\mathbf{S}_{\hat{\mathbf{j}}\omega}^{-r}\mathbf{S}_{\hat{\mathbf{j}}\omega}\mathbf{S}_{\hat{\mathbf{j}}\omega}^{-r} = \mathbf{S}_{\hat{\mathbf{j}}\omega}^{-r} \quad \left[based \; on \; \mathbf{K}^T\mathbf{S}_{\Phi\omega}^{-1}\mathbf{KTS}_{\Phi\omega}\mathbf{T}^T\mathbf{K}^T\mathbf{S}_{\Phi\omega}^{-1}\mathbf{K} = \mathbf{K}^T\mathbf{S}_{\Phi\omega}^{-1}\mathbf{K}\right]$

Eq. 15 defines the *partial covariance field*. And the *partial coherence field* is obtained by standardization of Eq. 15 as described in Eq. 13 and Eq. 14.

Note that Eq. 15 requires the existence of the inverse of the EEG covariance. If such an inverse does not exist because, e.g., the number of EEG epochs is smaller than the number of electrodes (i.e. $N_S < N_E$), we propose the use of a generalized inverse for the EEG covariance, thus giving the estimator for the inverse source covariance as:

**Eq. 18** $\mathbf{S}_{\hat{\mathbf{j}}\omega}^{-r} = \left(\mathbf{TS}_{\Phi\omega}\mathbf{T}^T\right)^{-r} = \mathbf{K}^T\mathbf{S}_{\Phi\omega}^+\mathbf{K}$

where the superscript "+" denotes the particular generalized inverse given by the Moore-Penrose inverse.

It should be noted that this estimator (Eq. 15) is based on the reflexive inverse, and not on the Moore-Penrose inverse, which would need to satisfy additional conditions, one of which is:

**Eq. 19** $\left(\mathbf{S}_{\hat{\mathbf{j}}_\omega}\mathbf{S}_{\hat{\mathbf{j}}_\omega}^{-r}\right)^* = \mathbf{S}_{\hat{\mathbf{j}}_\omega}\mathbf{S}_{\hat{\mathbf{j}}_\omega}^{-r}$

Using Eq. 7 and Eq. 15, this gives the requirement that:

**Eq. 20** $\mathbf{K}^T\mathbf{T}^T = \mathbf{TK}$

However, Eq. 20 is not satisfied in general. One example suffices for the proof, and easily given by any non-identity weighted inverse solution:

**Eq. 21** $\mathbf{T} = \mathbf{WK}^T\left(\mathbf{KWK}^T\right)^{-1}$

for $\mathbf{W} \neq \mathbf{I}$.

### 3.3. Essential properties of the inverse source covariance estimator

#### 3.3.1. The partial coherence field estimator is independent of the choice of linear inverse solution

As seen from Eq. 15, a first property is that the inverse source covariance is independent of any particular inverse solution, i.e., it is independent of any choice of $\mathbf{T}$ in Eq. 3 that satisfies Eq. 4.

#### 3.3.2. The partial coherence estimator for a particular pair of cortical voxels is explicitly independent of all other voxels

As seen from Eq. 15, a second property is that for any given pair of voxels, their inverse covariance, which is equivalent to their partial covariance after having accounted for the effect of all other voxels, is explicitly independent of all other voxels.

In formal terms, consider the voxel pair indexed by $(k,l)$. Let $\mathbf{k}_i \in \mathbb{R}^{N_E \times 1}$ denote the $i$-th column of the lead field matrix in Eq. 1. Then, from Eq. 15, the inverse covariance (i.e. partial covariance) of voxels $(k,l)$ is:

**Eq. 22** $\left[\mathbf{S}_{\hat{\mathbf{j}}_\omega}^{-r}\right]_{kl} = \mathbf{k}_k^T \mathbf{S}_{\Phi_\omega}^{-1} \mathbf{k}_l$

and the partial coherence is:

**Eq. 23** $\left[\mathbf{P}_{\hat{\mathbf{j}}_\omega}^{-r}\right]_{kl} = \frac{\mathbf{k}_k^T \mathbf{S}_{\Phi_\omega}^{-1} \mathbf{k}_l}{\sqrt{\left(\mathbf{k}_k^T \mathbf{S}_{\Phi_\omega}^{-1} \mathbf{k}_k\right)\left(\mathbf{k}_l^T \mathbf{S}_{\Phi_\omega}^{-1} \mathbf{k}_l\right)}}$

which are independent of the lead fields at other voxels.

This remarkable second property allows for very efficient computations of the inverse source covariance, without the need of an actual inversion of the enormous source covariance matrix in Eq. 7.

It is important to note that the inverse source covariance implicitly depends on the whole cortex covariance via the inverse of the EEG covariance.

#### 3.3.3. The resolution of the partial covariance field estimator

Let $\mathbf{S}_{\mathbf{J}_\omega} \in \mathbb{C}^{N_V \times N_V}$ denote the true source covariance, which will be assumed to be positive definite. Then, from Eq. 2, the true EEG covariance is:

**Eq. 24** $\mathbf{S}_{\Phi\omega} = \mathbf{K}\mathbf{S}_{\mathbf{J}\omega}\mathbf{K}^T$

and the estimated source covariance (in general, for any **T**) from Eq. 7 is:

**Eq. 25** $\mathbf{S}_{\hat{\mathbf{j}}\omega} = \mathbf{T}\mathbf{S}_{\Phi\omega}\mathbf{T}^T$

Using Eq. 24, the estimated reflexive generalized inverse of Eq. 25 is:

**Eq. 26** $\mathbf{S}_{\hat{\mathbf{j}}\omega}^{-r} = \left(\mathbf{T}\mathbf{K}\mathbf{S}_{\mathbf{J}\omega}\mathbf{K}^T\mathbf{T}^T\right)^{-r}$

which is independent of the choice of the inverse solution **T**.

Since any choice of **T** must give the same solution, we may choose the simplest Moore-Penrose (minimum norm) inverse solution:

**Eq. 27** $\mathbf{T}_{MinNorm} = \mathbf{K}^T\left(\mathbf{K}\mathbf{K}^T\right)^{-1}$

which gives:

**Eq. 28** $\mathbf{S}_{\hat{\mathbf{j}}\omega}^{-r} = \left(\mathbf{H}\mathbf{S}_{\mathbf{J}\omega}\mathbf{H}\right)^{-r}$

which is independent of the choice of the inverse solution **T**, with:

**Eq. 29** $\mathbf{H} = \mathbf{K}^T\left(\mathbf{K}\mathbf{K}^T\right)^{-1}\mathbf{K}$

Note that from Eq. 29, **H** is an idempotent projection matrix, and corresponds to the well-known resolution matrix of the minimum norm inverse solution.

Eq. 28 demonstrates the relation between the estimated inverse source covariance $\mathbf{S}_{\hat{\mathbf{j}}\omega}^{-r}$ and the actual source covariance $\mathbf{S}_{\mathbf{J}\omega}$. The estimator $\mathbf{S}_{\hat{\mathbf{j}}\omega}^{-r}$ is a function of the spatially filtered covariance, as "seen" through the filter given by the resolution matrix **H** (Eq. 29), which is known to have the effect of low spatial resolution.

## 4. An algorithm for the whole cortex electromagnetic connectivity

Step 0: Given an EEG Hermitian covariance matrix $\mathbf{S}_{\Phi\omega}$ (e.g. as in Eq. 8). Given the lead field matrix **K** (e.g. as used in Eq. 1).

Step 1: Compute the singular value decomposition of the EEG covariance matrix ($N_E \times N_E$):

**Eq. 30** $\mathbf{S}_{\Phi\omega} = \Gamma\Lambda\Gamma^*$

with $\Gamma$ denoting the eigenvectors, and the diagonal $\Lambda$ the non-zero eigenvalues.

Step 3: Compute the matrix **U**, corresponding to the Hermitian square root inverse of the EEG covariance:

**Eq. 31** $\mathbf{U} = \Gamma\Lambda^{-1/2}\Gamma^*$

NOTE: other inverse square root choices are possible.

Step 4: Compute:

**Eq. 32** $\mathbf{V} = \mathbf{K}^T\mathbf{U}$

and normalize each row. This can be expressed as:

**Eq. 33** $\mathbf{W} = \mathbf{E}\mathbf{V}$

with:

**Eq. 34** $\mathbf{E} = \left[ diag\left(\mathbf{VV}^T\right) \right]^{-1/2}$

NOTE: The matrices **V** and **W** are of size ($N_V \times N_E$); and in Eq. 34, only the diagonal elements of $\left(\mathbf{VV}^T\right)$ are needed, not the full matrix.

Step 5: The whole-cortex partial coherence matrix is:

**Eq. 35** $\mathbf{P} = \mathbf{WW}^T$

## 5. Some computational notes

These equations show that all the relevant connectivity information is contained in the low-rank square root matrix **W** of dimension $N_V \times N_E$ defined in Eq. 33. For instance, the largest left eigenvector can be used to compute distributed connections using the methodology and interpretation of Worsley et al (2005).

These equations are especially efficient when the aim is to view the whole cortex connections to a single point. For instance: how does the right auditory cortex connect to the rest of the cortex? In this case, the whole brain connectivities correspond to a single row (or column) of the full connectivity matrix Eq. 35, which is very easy to compute.

## 6. Results

A very simple simulation test was performed. Two cortical point sources were considered:

$x_t$: on cortex under scalp electrode Fp1, left frontal
$y_t$: on cortex under scalp electrode O2, right occipital

The time series for these true cortical current densities were generated as:

**Eq. 36** $\begin{cases} x_t = \varepsilon_{x,t} \\ y_t = 0.5 x_{t-1} + \varepsilon_{y,t} \end{cases}$

where both noise series $\varepsilon$ are independent and identically distributed uniformly in the interval $(-0.15...+0.15)$, i.e. $IID \sim U(-0.15...+0.15)$.

Additive biological noise, with 57 cortical neurons firing as $IID \sim U(-0.05...+0.05)$, was superimposed on the signals defined Eq. 36. Given such cortical current densities, the scalp potentials (EEG) were computed from Eq. 1 on 19 electrodes ($N_E = 19$) corresponding to the 10/20 system locations. Another layer of noise was added, consisting of measurement noise for the scalp potentials, as $IID \sim U(-0.05...+0.05)$ at each moment in time and at each electrode.

Using this procedure, 100 EEG epochs were generated ($N_S = 100$), each one consisting of 64 discrete time samples ($N_T = 64$). Assuming a sampling rate of 64 Hz, the EEG cross-spectrum was computed (see Eq. 8). Finally, using the EEG cross-spectral matrix for the alpha band (8-12 Hz), the classical (Eq. 9) and the new (Eq. 30 to Eq. 35) connectivities were computed, on 6239 cortical grey matter voxels ($N_V = 6239$).

In the figures, only "lagged classical" and "lagged partial" connectivities (as defined in Pascual-Marqui 2007; and Pascual-Marqui et al 2011) are displayed.

Define a "3D seeded connectivity map" as the connectivity values between a given cortical seed point with all other cortical voxels.

For illustration purposes, 19 seed points were used, consisting of the cortex underlying each electrode of the 10/20 system. The figures show the maximum connectivity at each voxel, over the set of 19 seeded connectivity maps. Figures 1 and 2 show the results for classical and partial connectivities, respectively.

The color scales of the images are proportionally identical, in such a way that the scale's maximum value was adjusted to 95% of the extreme connectivity value, thus allowing for a fair, unbiased comparison of the two competing methods.

For both types of connectivity measures (lagged classical and lagged partial), the maximum connectivities are in frontal and posterior regions. However, classical connectivity is extremely low resolution, with actual maxima not exactly located under Fp1 and O2. In contrast, and despite such noisy data and so few electrodes, the new partial connectivity measures have exact localization with very high resolution.

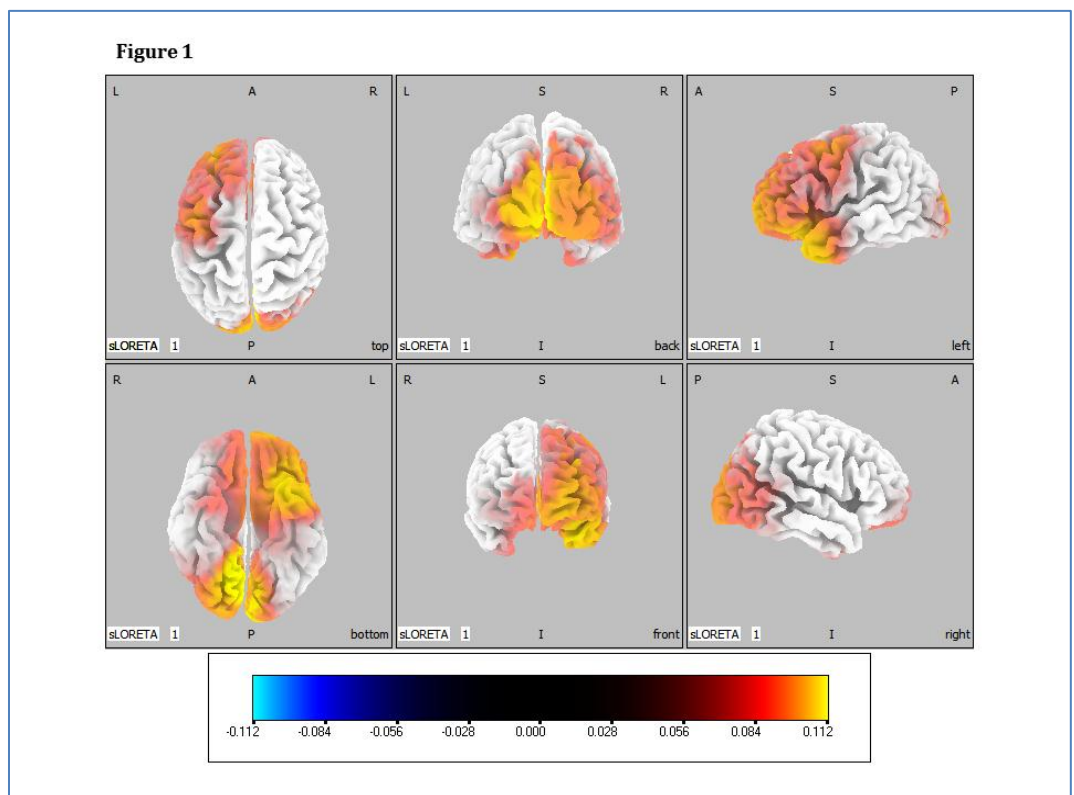

Figure 1: Classical connectivity map. L=left; R=right; A=anterior; P=posterior; I=inferior; S=superior. Actual connections correspond only to left frontal with right occipital.

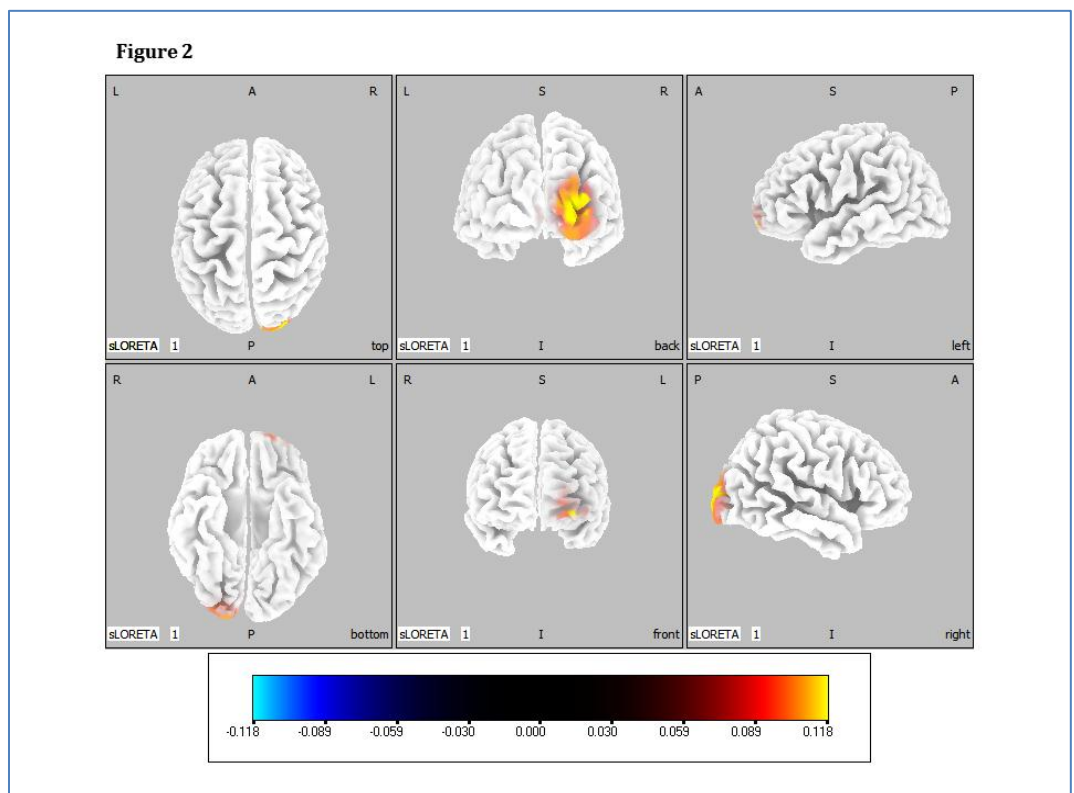

Figure 2: New partial connectivity map. L=left; R=right; A=anterior; P=posterior; I=inferior; S=superior. Actual connections correspond only to left frontal with right occipital.

## 7. Concluding remarks

Whole cortex connectivities can be estimated non-invasively with EEG and MEG. The new method uses a generalized reflexive inverse for the source covariance, which is endowed with the property of being invariant to any type of linear tomography. The method is very simple and efficient from a computational point of view.

The simulation results, based on only 19 EEG electrodes, with significant corruption by both biological and measurement noise, empirically show that the new method has much higher resolution than the classical connectivity measures.

The new method can be extended to time domain data, including the case of lagged partial correlation fields, thus providing information akin to Granger-type causality. Moreover, the general application field of this new method reaches beyond neuroimaging of connectivity fields.